\documentclass[]{aa}
\usepackage{natbib}
\usepackage[utf8]{inputenc}
\usepackage{tikz, graphicx}
\usepackage{subcaption}
\usepackage{txfonts}
\usepackage{color}
\usepackage{amsmath}
\usepackage{comment}
\usepackage{booktabs, tabularx}
\usepackage{placeins}
\usepackage{afterpage}

\graphicspath{{images/}}

\newcommand{\unit}[1]{\,\mathrm{#1}}
\newcommand{\msol}[1]{\,#1\,\mathrm{M}_\odot}

\definecolor{gaolvDarkGray}{gray}{0.5}
\newcommand{\skeleton}[1]{%
\ifx\showskeleton\undefined
\else
\vspace{1em}\noindent$\bigstar${\color{gaolvDarkGray}\emph{#1}}\newline%
\fi
}

\bibpunct{(}{)}{;}{a}{}{,} 

\let\linenumbers\relax 

\begin{document}
	\title{Gravitational trapping and ram pressure trapping \\ of ultracompact and hypercompact H II regions}
	\titlerunning{Modeling trapping of compact H~II regions}
	\author{
    L.~Martini \inst{1}
	\and
	A.~Oliva \inst{1,2,3}
	\and
	R.~Kuiper \inst{1}
    \and
    P.~Klaassen \inst{4}
	}
	\institute{
    Faculty of Physics, University of Duisburg-Essen, Lotharstraße 1, D-47057 Duisburg, Germany\\
    \email{rolf.kuiper@uni-due.de}
    \and
    Département d'Astronomie, Université de Genève, Chemin Pegasi 51, 1290 Versoix, Switzerland
    \and
    Space Research Center (CINESPA), School of Physics, University of Costa Rica, 11501 San José, Costa Rica
    \and
    UK Astronomy Technology Centre, Royal Observatory Edinburgh, Blackford Hill, Edinburgh EH9 3HJ, United Kingdom
	}


	\abstract{
    Observationally, early H~II regions are classified by size into ultracompact and hypercompact configurations. It remains unclear whether these phases are long-lived or transient. Understanding the physical processes that stall H~II region growth may help to solve the ``lifetime problem'': the observation of more compact H~II regions than expected from theory.
	}{
 	Utilizing two-dimensional, axially symmetric radiation hydrodynamic simulations of young expanding H~II regions, including the phase of early star and disk formation, we seek to better understand the trapping of H~II regions. Trapping forces include gravity and ram pressure, which oppose forces such as thermal pressure expansion, radiation pressure, and centrifugal force.
	}{
	We use the open-source MHD code \emph{Pluto}, the radiation transport module \emph{Makemake}, the photoionization module \emph{Sedna} and the self-gravity module \emph{Haumea}. 
    This is similar to \citet{kuiper2018first} without manual injection of protostellar outflows.
	}{
	Without radiation pressure, the H~II region remains gravitationally trapped in the ultracompact phase indefinitely. With radiation pressure, the H~II region escapes gravitational trapping but experiences ram pressure trapping on larger scales.
    For initial mass reservoirs with high central density, no trapping occurs, while a less steep density gradient yields clear trapped phases. Hypercompact trapped phases exhibit a ``flickering'' variation in H~II region radius, in agreement with observations of stalling and even contraction over small time scales.
    With radiation pressure, low-mass/low-density reservoirs experience both gravitational and ram pressure trapping, while high-mass reservoirs undergo only the latter.
	}{
    During the early evolution of H~II regions, gravitational and ram pressure trapping may occur, at different times and distances from the (proto)star, depending on environmental conditions. The mean expansion velocity is influenced by thermal pressure, gravity, centrifugal forces, and radiation pressure. For the cases studied herein, the inclusion of radiation pressure is crucial to obtain physically meaningful results.
	}

	\keywords{HII regions, Hydrodynamics, Methods: numerical, Stars: massive, Stars: formation} 

	\maketitle

\section{Introduction}
During the formation of massive stars, the extreme ultraviolet (EUV) radiation of the protostar may ionize its surroundings. This ionized region is known as an H~II region. The expansion of this region is a key form of stellar feedback, as H~II regions make up a significant component of the interstellar medium (ISM). 
They offer an observational signpost for the otherwise difficult-to-observe forming massive stars, as this is the most direct evidence that a massive star has formed and has begun to ionize its environment. If an H~II region is observed, there must be at least one massive star responsible for it.
Though H~II regions can grow to hundreds of parsecs in diameter \citep{oey2003h}, our interest lies in studying them at their smallest, during the early stages of star formation. 
Young H~II regions fall roughly into the following categories based on radius:
hypercompact (HC-) for 0.001 to 0.01$\unit{pc}$, ultracompact (UC-) for 0.01 to 0.1$\unit{pc}$, and compact for 0.1 to 0.5$\unit{pc}$. These are the ranges used in works such as \citet{gaume1995ngc}, but note that other works may choose to associate different radius ranges with the same terms \citep[see e.g.][]{kurtz2005hypercompact}. These size categories are primarily observational, based on the resolved angular size of the regions \citep{gaume1995ngc}, rather than being tied to distinct physical transitions.

\subsection{Observations of ultra- and hypercompact H~II regions}
The first survey of UCH~II regions was presented in \citet{wood1989morphologies}, in which 75 sources were identified. Prior to 1989, only a handful of UCH~II regions had been described. \citet{gaume1995ngc} presented the first HCH~II region, prompting further surveys for H~II regions in this size range in the following decades.

By the mid-2000s, eight major surveys of UCH~II regions had been performed, resulting in hundreds of identifications, while a handful of HCH~II regions had been found. See \citet{kurtz2005hypercompact} for a summary of these surveys. In the 2010s and beyond, UCH~II and HCH~II surveys included \citet{urquhart2013atlasgal} based on the ATLASGAL survey, \citet{cesaroni2015infrared} based on the Herschel/Hi-GAL survey, \citet{kalcheva2018coordinated} based on the CORNISH survey, \citet{klaassen2018evolution} which presented ALMA observations of 9 young H~II regions, and \citet{yang2021population} which identified 16 HCH~II candidates from VLA observations. Observations of individual UCH~II/HCH~II regions include \citet{beuther2022feedback}, \citet{cesaroni2019depth}, \citet{maud2018chasing}, \citet{moscadelli2018feedback}, \citet{beltran2018accelerating}, \citet{klaassen2009rotation}, \citet{sollins2005spherical}, \citet{keto2002ionized}, \citet{ball1996thermal}, and \citet{guzman2014disk} among many others.

\subsection{Physics of expansion and morphology} 
\skeleton{Gravitational trapping}

H~II region expansion has two phases, each driven by a different physical process: R-type (short for ``Rarefied'') and D-type (short for ``Dense''). The initial, supersonic R-type expansion phase is photoionization-driven and very short-lived. An ionization front, the boundary between the ionized and neutral hydrogen gas, forms and moves outwards from the protostar. This phase ends once the ionization front reaches the Str{\"o}mgren radius, at which there is equilibrium between ionization and recombination. Subsequently, the high thermal pressure of the hot (${\sim}10^4$ K) ionized gas drives slower, subsonic D-type expansion into the surrounding cold neutral medium \citep{kahn1954acceleration, draine2011physics}.

Once an H~II region begins pressure-driven D-type expansion, its theoretical expansion velocity is on the order of the sound speed of the ionized gas (${\sim}10$ km/s). At this speed, an H~II region would expand beyond the ultracompact phase in only (${\sim}10^4$ years) \citep{comeron1996galactic}. This short theoretical lifetime is in conflict with the large observed number of compact H~II regions, suggesting that one or more mechanisms must be acting to restrict or stall their expansion \citep{kurtz2000u}.

One possibility is that the ionized gas is initially prevented from beginning R-type expansion because it is gravitationally bound to the protostar. This mechanism is referred to as gravitational trapping. Gravitational collapse also leads to an inward flow of gas towards the protostar, which impacts the H~II region at larger radii later on. At these later stages during slower D-type expansion, the infalling gas slows H~II region expansion via ram pressure applied directly against the expanding ionized gas \citep{walmsley1995dense}. We refer to this mechanism as ram pressure trapping.

\skeleton{radiation pressure as a trapping-opposing feedback effect}
During both early and late expansion phases, continuum radiation pressure from the forming star contributes to pushing the (dusty) ionized gas outwards. 
Furthermore, line-driven stellar winds contribute to disk ablation and ending the accretion phase \citep{kee2019line}.

\begin{figure}[t]
    \centering
    \begin{subfigure}{0.35\textwidth} 
         \centering
         \includegraphics[width=\textwidth]{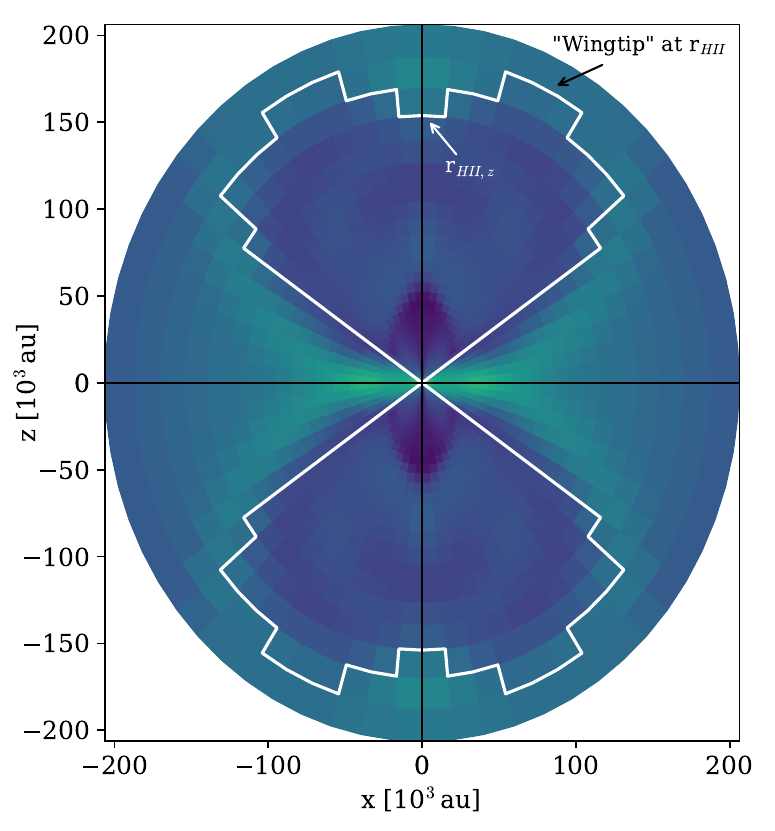}
     \end{subfigure}
     \begin{subfigure}{0.09\textwidth} 
         \centering
         \vspace{-6.5mm}
         \includegraphics[width=\textwidth, trim= 0 0.25cm 0 0cm, clip=true]{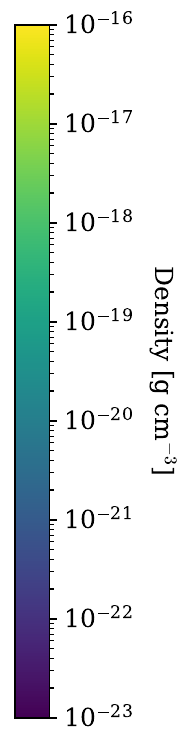}
     \end{subfigure}
     \caption{Example snapshot of typical H~II region ``butterfly wing'' morphology taken from run \#0 in this work, as a density color-map. The boundary depicted with the white outline is the region where more than 50\% of the hydrogen was ionized.}
    \label{fig:butterfly_morphology}
\end{figure}

\skeleton{centrifugal forces}
Additionally, the rotation of the gas applies a centrifugal force cylindrically outwards. This may guide the H~II region into a characteristic ``butterfly'' morphology, with top and bottom lobes around the central axis shaped on the midplane by an accretion disk. See Figure \ref{fig:butterfly_morphology} for a simulation snapshot with an H~II region demonstrating the classic ``butterfly'' morphology, and Figure \ref{fig:butterfly_expansion_snapshots} for the development of this morphology over time. Given that the centrifugal force is directed outwards, it may also contribute towards expansion. 

\skeleton{relative motion of the star, blister-type H~II region} 
The proper motion of the star itself through its surroundings can also change the morphology and expansion rate of the H~II region. 
If the ionizing star moves to the edge of a cloud, this can also result in ``champagne flows`` \citep{tenorio1979gas} and ``blister-type`` (cometary) shapes \citep{israel1978h}. 
Proper motion is not addressed further in this work, but see \citet{arthur2006hydrodynamics} for more information.

\subsection{The Lifetime Problem}
\label{Section-Lifetime Problem}
The ``Lifetime Problem`` refers to the aforementioned discrepancy between the theoretically predicted lifetime of H~II region compact phases and the observed ratio of compact H~II regions to non-compact. Assuming expansion to occur at sound speed unhindered by feedback, we expect to observe far fewer HCH~II, UCH~II, and compact H~II regions than we actually observe \citep{kurtz2000u}. As far back as the groundbreaking survey in \citet{wood1989morphologies}, UCH~II region lifetimes were determined to last even into the main sequence phase of stellar evolution ($10^{5} - 10^{6}\unit{years}$), based on the number of H~II regions observed in each phase. Later, \citet{comeron1996galactic} predicted lifetimes of $10^{4}\unit{years}$ with statistical analysis of UCH~II observations, but such lifetimes still require a mechanism to slow down expansion. \citet{mottram2011rms} estimated lifetimes closer to those in \citet{wood1989morphologies}.

The lifetime problem may be resolved if the expansion of compact H~II regions is not always monotonic. Both simulations and observations suggest that these regions can stall or even contract over years to decades. This ``flickering'' phenomenon is supported by multi-epoch observational studies that have reported significant variability in compact H~II region emissions. For example, \citet{franco2004time} reported a 20\%-30\% decrease in radio continuum flux density from the UCH~II region NGC 7538 IRS 1 over 11 years, while \citet{de2014flickering} and \citet{de2018flux} observed similar decreases in sources within the massive star-forming regions Sgr B2 and W49A, respectively. 

A theoretical basis for this flickering is provided by three-dimensional simulations from \citet{peters2010h}, which suggest that accretion flows onto massive stars are gravitationally unstable, creating dense clumps and filaments that cause the H~II region to rapidly fluctuate in size. Notably, these influential simulations did not include radiation pressure. While accretion-driven flickering provides a compelling solution to the lifetime problem, the role of radiation pressure in this process remains unclear. In this work, we use numerical simulations to investigate the effect of radiation pressure on the trapping and expansion, and thus on the lifetimes, of compact H~II regions.



\subsection{Motivation}
The basic aim of this numerical experiment is to check if the expansion of the H~II region during its D-type evolutionary phase can be halted for some time due to either the gravity of the hot host star and/or by the ram pressure from the in-falling gas from larger spatial scales. 
In the cases in which the H~II region is trapped, we measure the lifetime of these trapped phases and check for the corresponding sizes of the H~II region in comparison to the commonly used categories of ``hypercompact'' ($\le 0.01 \mbox{ pc} \sim 2000 \mbox{ au}$) and ``ultracompact'' ($\le 0.1 \mbox{ pc} \sim 20000 \mbox{ au}$) H~II regions. 
Furthermore, to check the impact of radiation forces and whether the trapping, the sizes, and/or the timescales depend on the mass distribution of the initial cloud reservoir, we perform simulations for different core masses and different density slopes and we run all of these setups once with and without radiation forces included.

\section{Methods}

\subsection{Physics}
This work describes two-dimensional, axially and midplane symmetric simulations. This configuration enables the modeling of circumstellar disk formation and evolution simultaneous with the earliest expansion of H~II regions \citep{kuiper2018first}. 

In order to study the viability and strength of the proposed trapping processes, we run direct numerical simulations which include all of the relevant physical processes:
\begin{itemize}
    \item Photoionization feedback from stellar irradiation allows us to model the R-type expansion phase of early H~II regions.
    \item The hydrodynamics evolution of the system allows us to study the subsequent D-type expansion phase.
    \item Gravity of the star and self-gravity of the surrounding gas is included to check for the viability of gravitational trapping.
    \item By following the evolution of the system from the large-scale collapse of a massive reservoir of gas and dust down to the formation of the EUV-emitting star, the models self-consistently include the effects of large-scale infall and ram pressure during the D-type expansion phase of the H~II region.
    \item The radiation transport modules of continuum radiation and photoionization both include the impact on the hydrodynamic evolution of the surrounding gas and dust via by radiation forces due to absorbed momentum \citep{kuiper2020makemake+}.
\end{itemize}
We also study the effect of different environments on these physical processes by varying the total mass of the initial reservoir as well as its initial density distribution.

\begin{figure}[htbp]
    \centering
    \includegraphics[width=0.41\textwidth]{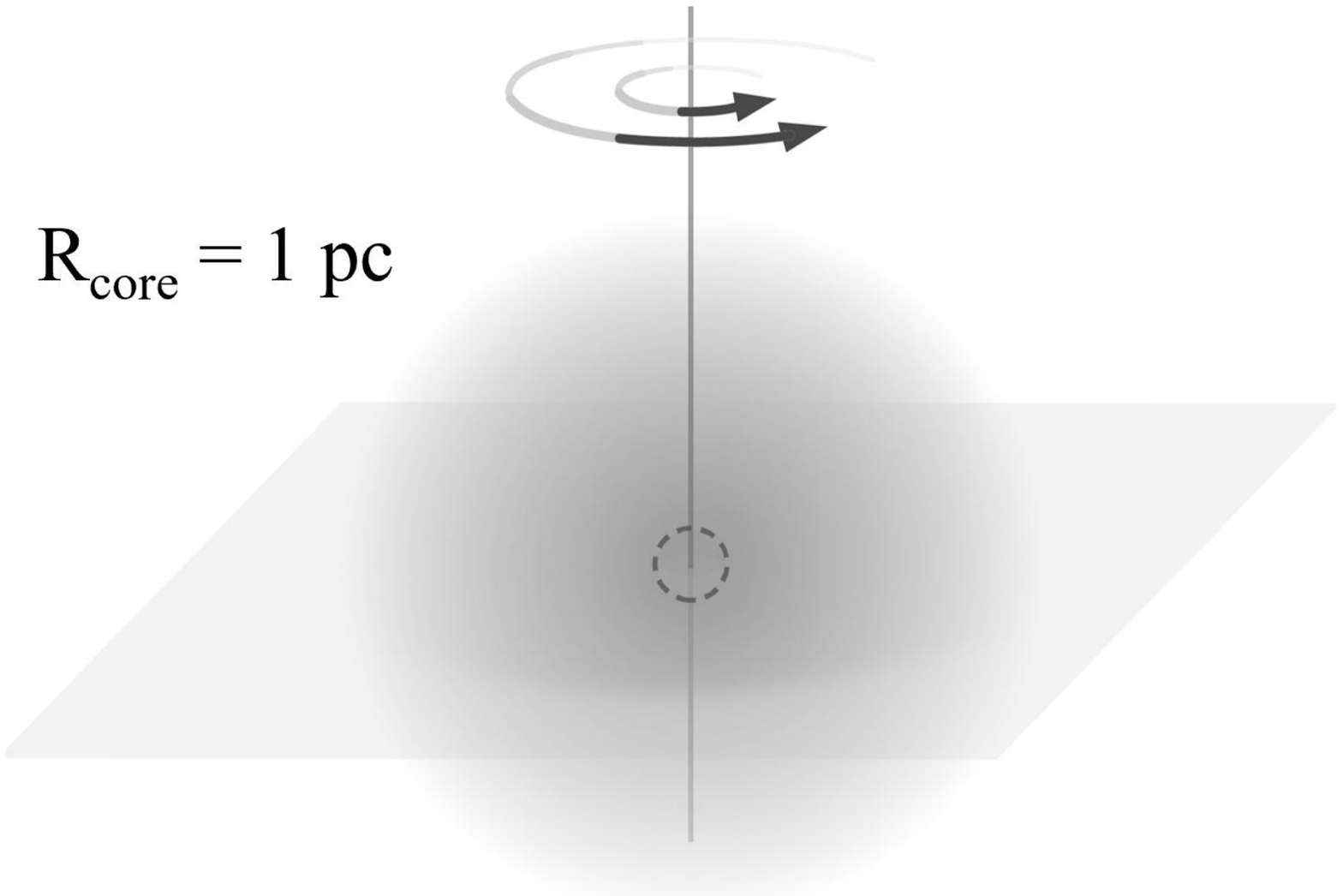}
    \caption{The cloud at the start of the simulation before collapse, displayed in 3D. Rotation and the midplane are indicated.}
    \label{fig:initial_cloud}
\end{figure}

We start with a spherical core of gas and dust which collapses under its own self-gravity, representing, for instance, a high-density region within a giant molecular cloud (GMC). 
See Figure \ref{fig:initial_cloud} for a visualization of the initial state of the mass reservoir. The setup of the simulations is similar to the one described in \citet{kuiper2018first} but without the injection of protostellar outflows. 
During the course of the simulation, a protostar with an accretion disk forms from the collapsing gas, then begins emitting both from its interior and from the accretion shock front. The EUV component of these emissions ionizes hydrogen around the star, creating an H~II region.

\skeleton{Hydrodynamics}
For the hydrodynamics of the gas and dust, we use the open-source grid-based code Pluto \citep{mignone2007pluto, mignone2012conservative}. We include an $\alpha$-shear viscosity prescription for modeling the angular momentum transport in the accretion disk due to gravitational torques. The system is modeled as a calorically perfect gas, with additional acceleration source terms for self-gravity, the gravity of the forming massive star, radiation transport, and photoionization. 

\skeleton{Gravity}
The self-gravity solver \emph{Haumea} is described in more detail in \citet{kuiper2010circumventing}. We also reference the gravitational binding radius given by \citep{keto2003formation}:
\begin{equation*}
\label{eq:binding}
    r_{\text{binding}} = \frac{G M_{*}}{2c_{\text{s,ion}}^{2}}
\end{equation*}
where $c_\text{s,ion}$ is the isothermal sound speed of the ionized gas. This is the theoretical radius at which the thermal pressure expansion of the hot H~II region is balanced by the gravitational pull of the central stellar mass $M_{*}$.

\skeleton{Radiation transport, stellar evolution}
Radiation pressure is derived from the radiation field, which is calculated with the module \emph{Makemake} \citep{kuiper2020makemake+} using the two-temperature approach. In short, this module divides the total radiation flux into two components: frequency-dependent stellar irradiation and diffusive thermal (re-)emission by dust and gas. The irradiation flux is computed via ray tracing, for which we use the dust opacities from \citet{laor1993spectroscopic}. More specifically, the spectrum of the forming massive star is divided into 56 frequency bins. The non-ionizing part of the spectrum is resolved by 54 of these bins, which are used to accurately determine the dust heating and the radiation force via direct stellar irradiation. The FUV regime ($6\,\mathrm{eV} \leq h\nu \leq 13.6\,\mathrm{eV}$) and the EUV regime ($h\nu \ge 13.6 \mathrm{\,eV}$) are each treated as a single bin for the purpose of transport. While using a single bin for the EUV radiation is an approximation, the total energy in this bin is calculated by first filtering the stellar spectrum with a realistic stellar atmosphere model to ensure an accurate ionizing flux \citep{kuiper2020makemake+}. 

The effective absorption cross-section for the single EUV bin is not constant, but is calculated as a function of the stellar temperature ($T_{star}$) using the analytical fit: 
\[\sigma_{EUV} = (21.6 - 4 \log(T_{star}/\mathrm{K})) \times 10^{-18} \mathrm{cm^2}\]
See \citet{kuiper2020makemake+} for further detail. For a typical stellar temperature of 40,000 K, this results in a cross-section of approximately $3.2 \times 10^{-18} \mathrm{\,cm^2}$.

The dust opacities for the FUV and EUV bins are $\kappa_\text{dust} = 4\times 10^4 \mathrm{\,cm^2\,g^{-1}}$ and $2\times 10^4 \mathrm{\,cm^2\,g^{-1}}$, respectively. The gas has a constant opacity of $0.01\,\mathrm{cm^2\,g^{-1}}$. 

The thermal (re-)emission flux is treated as frequency-independent, and computed with the flux-limited diffusion approximation. For protostellar evolution, we use the evolutionary tracks by \citet{hosokawa2009evolution}.

\skeleton{Ionization}
Hydrogen photoionization is handled by the module \emph{Sedna} \citep{kuiper2020makemake+}, which we summarize as follows. The part of the spectrum of the proto(star) with photons capable of ionizing hydrogen is first filtered, also taking into account the effect of a Kurucz-like stellar atmosphere instead of a simple black-body spectrum, and ray tracing is performed in order to compute the radiative EUV flux:
\[ F_\text{EUV}(r) = F_\text{EUV}(r_\text{min}) \left(\frac{r_\text{min}}{r}\right)^{2} \exp(-\tau_\text{tot})  \]

Here, $r_\text{min}$ corresponds to the radius of the central sink cell ($3\unit{au}$), and the initial flux, $F_\text{EUV}(r_\text{min})$, is determined from the stellar evolution module based on the star's current mass and radius. The factor $\exp(-\tau_\text{tot})$ denotes the total extinction of the radiative flux along the path, and the optical depth $\tau_\text{tot}$ has contributions from the ionization of hydrogen and the continuum extinction by dust grains and gas. The contribution to the optical depth due to ionization depends on the ionization fraction $x$, which is computed simultaneously via rate equations, which express how $x$ changes in a given region due to radiative ionization, collisional excitation, and recombination.

To handle the net effect of recombination, we use the on-the-spot approximation, which assumes that all diffuse Lyman continuum photons produced by recombinations to the ground state are immediately re-absorbed locally.

The gas temperature is set by a physical prescription based on the local ionization fraction, which is a valid approach for photoionized regions where the temperature is set by ionization and recombination equilibrium rather than hydrodynamic heating or cooling. The gas temperature of a fully ionized region is set to $8000\,\mathrm{K}$, while the temperature of a fully neutral medium is assumed to match that of the dust. The temperature of a partially ionized medium is computed as a linear interpolation between these states. This temperature is then used to calculate the gas pressure, which provides the back-reaction on the hydrodynamic momentum and continuity equations.

To quantify the radiative feedback from the central star, we calculate the ionizing photon rate ($S_*$) at each timestep. This rate is derived from the ionization-irradiation temperature ($T_{ion}$) and radius ($r_\text{ion}$) provided by the stellar evolution module. The photon rate is determined by calculating the total ionizing luminosity ($L_{ion}$) and dividing by the average energy per ionizing photon. Given an average photon energy equal to the ionization potential of hydrogen ($13.6\mathrm{\,eV}$), the rate is given by:\\
\[ S_* = 4 \pi R_{ion}^2 \sigma T_{ion}^4 \,/\, 13.6\mathrm{\,eV} \]

\skeleton{Omission of Outflows}
We generally follow the setup described in \citet{kuiper2018first} but disable the manual injection of early protostellar outflows. In that work, injected protostellar outflows created large, low-density bipolar cavities, allowing the H~II region to expand freely without trapping. This outflow-driven simulation scenario in \citet{kuiper2018first} provides a model for observational counterparts such as S106 \citep{bally1983radio}. In this study, we investigate the opposite limit by disabling the manual injection of outflows to determine if trapping is a natural occurrence in their absence. 
Another possible application scenario would be those cases where the occurrence of protostellar outflows and the expansion of H II regions are separated in time. 

\subsection{Geometry and boundary conditions}
The mass reservoir has a spherical shape and a size of $1\mathrm{\,pc}$. We use an axisymmetric, two-dimensional grid in spherical coordinates that also assumes midplane symmetry. There is a sink cell at the center of the mass reservoir of $3\unit{au}$ in radius; we refer the reader to \citet{kuiper2018first} for a convergence study on the size of the sink cell. The massive (proto)star is formed inside of the sink cell during the simulation. Matter is allowed to flow into the sink cell, but not to leave it. There are 114 cells in the radial direction, scaling logarithmically with distance to the sink cell. Matter may exit at the outer boundary but not re-enter. The polar angle uses 16 cells and extends uniformly from $\theta=0$ to $\theta=\pi/2$. The size of the mass reservoir ($1\unit{pc}$) is large enough so that its free-fall time is much longer than the accretion time of the forming star, so that it can be thought as virtually infinite.

\subsection{Initial conditions}
The initial rotation is given by an angular velocity profile inversely proportional to the distance $R$ from the rotation axis: $\Omega \propto R^{-1}$. The proportionality constant is fixed by the initial rotation to gravitational energy ratio, set to 2\%. 
The gas temperature is initially uniform, and set to $10\unit{K}$.
The initial density profile is spherically symmetric with the following dependence on distance $r$ from the origin:
\[\rho(r) = {\rho}_0 \left(\frac{r}{R_\text{core}}\right)^{\beta_\rho},\ \ {\rho}_0 = \frac{M_\mathrm{core} \times (\beta_\rho + 3)}{4\pi R_\text{core}^3} \]
where $R_\text{core}$ is the radius of the cloud. 

Two initial density profiles are used: one with $\beta_{\rho} = -1.5$ and one where $\beta_{\rho} = -2.0$. In the latter case, the cloud mass is more concentrated closer to the star. 
A setup with $\beta_{\rho} = -2.0$ yields a higher accretion rate, faster initial growth of the star, and a higher accretion luminosity feedback, while a setup with $\beta_{\rho} = -1.5$ yields stronger ram pressure from the infall of the gas from large scales. 
The exact impact of the different reservoirs masses and density profiles is further complicated by the fact that the accretion histories lead to different durations of the bloating phase of the protostar, which delays the onset of photoionization feedback \citep{kuiper2018first}.
The two values of the density profiles used mark the most extreme cases for observed density profiles in high-mass star forming regions on scales up to approximately $0.1 \mbox{ pc}$ to $1 \mbox{ pc}$ \citep{gieser2022clustered, beuther2024density}.
See Table \ref{tab:parameter_summary} for all combinations of $M_\text{core}$ and $\beta_\rho$ used.
\begin{table}[tb]
    \caption{Simulation parameter summary}
    \label{tab:parameter_summary}
    \centering
    \begin{tabular}{r r r r r}
        \hline\hline
        Run \# & Run ID & $M_\mathrm{core} \mbox{ [M}_\odot]$ & $\beta_\rho$ & $P_{rad}$\\
        \hline
        0 & 250-1.5-RAD & 250 & -1.5 & On\\
        1 & 250-2.0-RAD & 250 & -2.0 & On\\
        2 & 500-1.5-RAD & 500 & -1.5 & On\\
        3 & 500-2.0-RAD & 500 & -2.0 & On\\
        4 & 1000-1.5-RAD & 1000 & -1.5 & On\\
        5 & 250-1.5-NORAD & 250 & -1.5 & Off\\
        6 & 500-1.5-NORAD & 500 & -1.5 & Off\\
        7 & 1000-1.5-NORAD & 1000 & -1.5 & Off\\
    \hline
    \end{tabular}
\end{table}

\section{Results and Discussion}

We performed eight simulations in two sets. The first five simulations (runs \#0-4) included radiation pressure, while the latter three simulations (runs \#5-7) had radiation pressure switched off. For now, we consider the simulations with the initial density profile $\beta_\rho=-1.5$. Table \ref{tab:parameter_summary} contains all parameter combinations used. Figure \ref{fig:rad_pressure_comparison} displays the growth of the H~II region radius ($r_\text{HII}$) as the (proto)star increases in mass, for each of the initial reservoir mass conditions and with radiation pressure switched on and off.

\subsection{Gravitational trapping and radiation pressure}

\begin{figure}
    \includegraphics[width=0.49\textwidth]{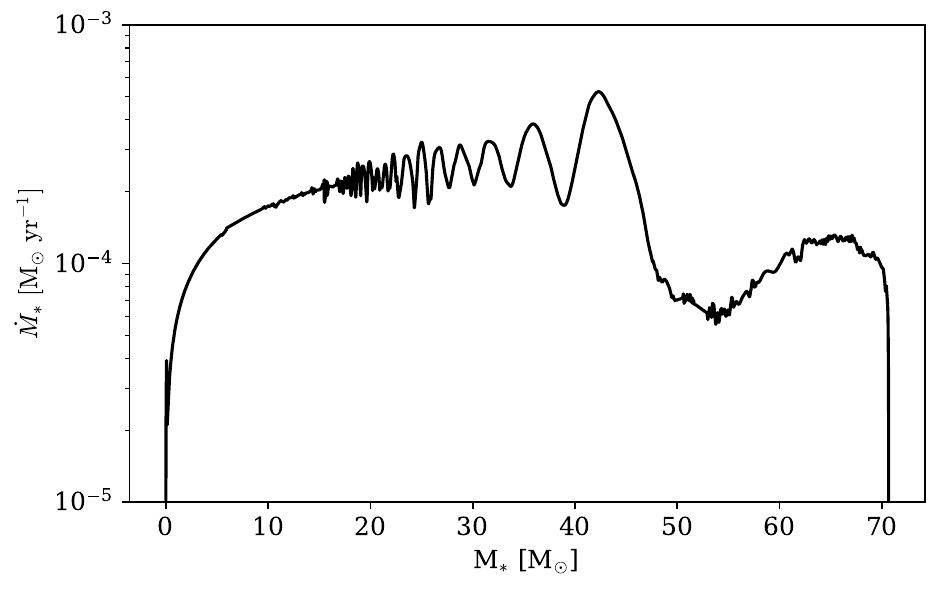}
    \caption{Accretion rate ($\dot{\mathrm{M}}_{*}$) onto the protostar as a function of stellar mass for the representative $\msol{500}$ simulation (run \#2).} \label{fig:accretion_rate_vs_stellar_mass}
\end{figure}

\begin{figure}
    \includegraphics[width=0.49\textwidth]{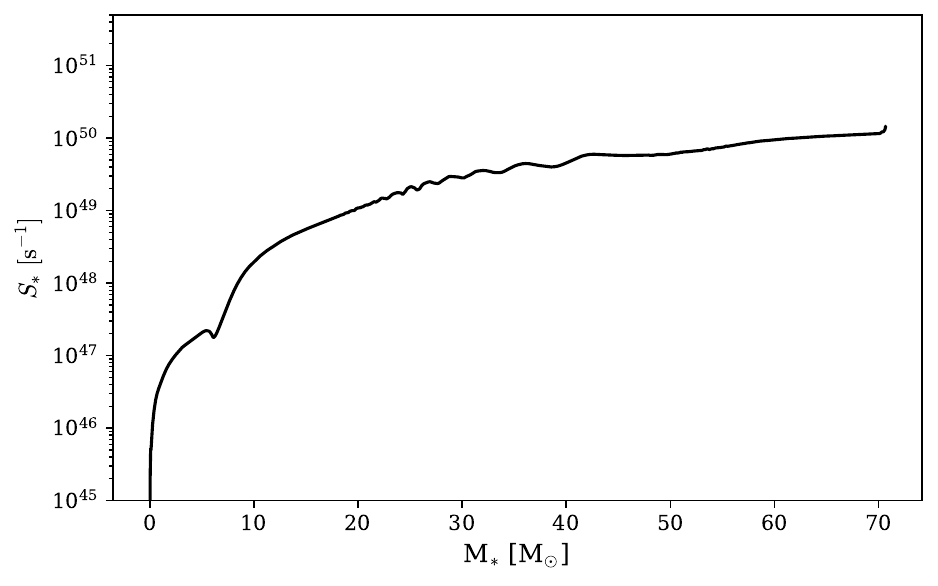}
    \caption{The ionizing photon rate ($S_*$) as a function of stellar mass for the representative $\msol{500}$ simulation (run \#2).}  \label{fig:ionizing_photon_rate_vs_stellar_mass}
\end{figure}

\begin{figure*}[htbp]
    \centering
    \includegraphics[width=\textwidth]{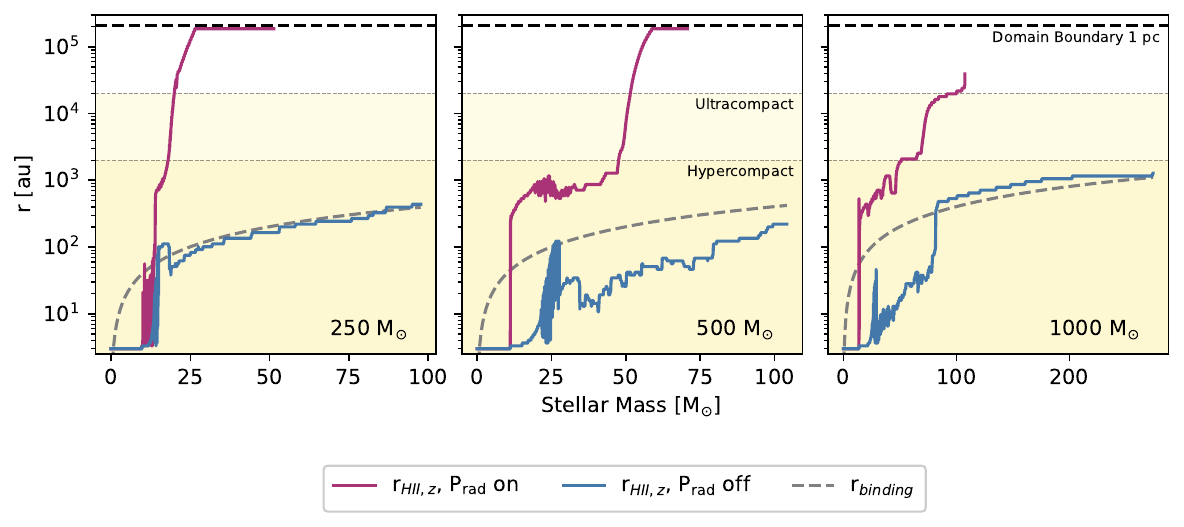}
    \caption{From left to right: H~II region extent with (magenta) and without (blue) radiation pressure switched on, in the $M_{\text{core}} = \msol{250}$ (runs \#0, \#5), $\msol{500}$ (runs \#2, \#6) and $\msol{1000}$ (runs \#4, \#7) cases.}
    \label{fig:rad_pressure_comparison}
\end{figure*}

The first barrier the H~II region must surpass in order to expand is escaping the gravity of the (proto)star. Figure \ref{fig:rad_pressure_comparison} includes a representation of this barrier in the critical radius $r_\text{binding}$. In the absence of radiation pressure, the gravitational pull of the (proto)star was sufficient to keep the H~II region trapped for the duration of the simulation in all mass cases. 

Case 250-1.5-NORAD provides a clear example of gravitational trapping in the absence of radiation pressure, as the H~II region boundary $r_\text{HII, z}$ on the central axis cleanly and continually followed the gravitational binding radius. This coupling indicates gravitational trapping. The binding radius represents the equilibrium point where the outward thermal pressure of the ionized gas is balanced by the inward gravitational pull of the (proto)star. In the absence of other strong outward forces like radiation pressure, the H~II region's D-type expansion halts at this boundary. The agreement between simulated H~II region extent and $r_\text{binding}$ thus provides direct evidence for the gravitational trapping mechanism.

\begin{figure*}[htbp]
    \centering
    \includegraphics[width=\textwidth]{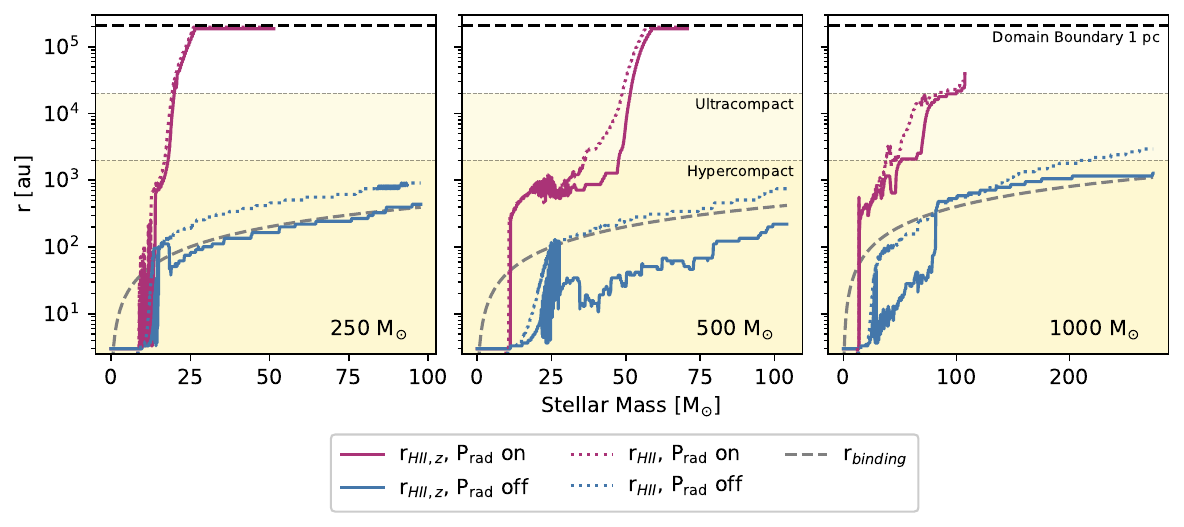}
    \caption{For the three initial mass cases and with radiation pressure on and off (same run numbers as in Figure \ref{fig:rad_pressure_comparison}), compare the H~II region radius on the central axis $r_\text{HII,z}$ (solid lines) with the maximum H~II region radius away from the central axis $r_\text{HII}$ (dotted lines).}
    \label{fig:rHIIz_rHII_comparison}
\end{figure*}

Away from the central axis, centrifugal force extended the boundary $r_\text{HII}$ further out. Figure \ref{fig:rHIIz_rHII_comparison} shows the difference between the boundary extent on the z-axis $r_\text{HII,z}$ and the maximum extent away from the z-axis $r_\text{HII}$. Also refer to Figure \ref{fig:butterfly_morphology}, as $r_\text{HII}$ corresponds to the tips of the ``wings'' as depicted in 2D. However, even this maximum boundary remained within the hypercompact phase for the entire simulation time. Similarly, in cases 500-1.5-NORAD and 1000-1.5-NORAD, the H~II region was unable to escape the gravity of the (proto)star and its surroundings. The H~II region in 500-1.5-NORAD failed to escape the hypercompact phase, and the maximum boundary radius away from the axis tracked $r_\text{binding}$ closely. In 1000-1.5-NORAD, the boundary at the axis followed $r_\text{binding}$, while away from the axis the boundary very gradually diverged from $r_\text{binding}$ until just barely escaping into the ultracompact phase late when the star was already over $\msol{200}$. In all cases without radiation pressure, the expansion of the H~II region was visibly gravitationally constrained for the simulated lifetime of the (proto)star.

In contrast, in the presence of radiation pressure, the gravitational binding radius was rapidly surpassed for all mass cases. 250-1.5-RAD exhibited a short, oscillatory gravitationally trapped phase from approximately $210$ to $268\unit{kyr}$ after the initial launching of the H~II region via rapid photoionization-driven R-type expansion. The subsequent expansion beyond this point is a pressure-driven D-type phase, which requires the additional outward force from radiation pressure to succeed, as we will establish in this work. 

At these early stages, the accretion luminosity dominates the total energy output, and fluctuations in the accretion rate (see the oscillatory period in Figure \ref{fig:accretion_rate_vs_stellar_mass}) lead to corresponding changes in the ionizing photon flux, causing the H~II region to expand and contract on short timescales. A sudden increase in the accretion rate can lead to a temporary drop in the ionizing photon flux reaching the outer parts of the H~II region, as the dense, infalling material absorbs photons close to the star. With the outward push from thermal and radiation pressure suddenly weakened, the constant inward forces of gravity and ram pressure temporarily dominate, causing the H~II region to contract into a trapped state.

In the higher mass cases, the H~II region quickly surpasses the binding radius, skipping the gravitational trapping phase seen in 250-1.5-RAD. This difference in trapping behavior arises from the different initial conditions of the models. The higher-mass runs begin with a higher average density, leading to a shorter collapse timescale and more rapid stellar evolution. Consequently, in these cases, the central star's radiative feedback overpowers the gravity before a trapped state can be established, allowing the H~II region to quickly surpass the binding radius. The ability to eventually overcome this trapping is directly linked to the increasing radiative feedback as the star grows. In Figure \ref{fig:ionizing_photon_rate_vs_stellar_mass}, the ionizing photon rate ($S_*$) in 500-1.5-RAD increases by several orders of magnitude, from approximately $10^{47} \mathrm{\ s^{-1}}$ when the star is $\msol{10}$ to over $10^{49} \mathrm{\ s^{-1}}$ by the time the star reaches $\msol{20}$. This increase in radiative feedback (photoionization feedback and radiation pressure) pushes the H~II region beyond its binding radius and drives the subsequent expansion.

The outwards boost from radiative feedback not only contributes to expansion, but is required to prevent indefinite trapping within the hypercompact phase. This would suggest that radiation pressure may not be omitted from simulations of early massive star formation without losing significant realism.
Moreover, H~II regions observed in the hypercompact phase might be gravitationally trapped due to diminished radiation pressure. 

\subsection{Ram pressure trapping}
While the gravitational trapping phase was short to non-existent in the presence of radiation pressure, an additional trapped phase occurred beyond $r_\text{binding}$ due to the ram pressure of infalling gas. In all three mass cases with radiation pressure, there was at least a short trapped phase due to ram pressure when $r_\text{HII}$ was approximately $1000\unit{au}$. These phases appear in Figure \ref{fig:rad_pressure_comparison} as interruptions in expansion beyond $r_\text{binding}$. 250-1.5-RAD experienced the least ram pressure trapping, with only a brief slowing of expansion at $268\unit{kyr}$ ($M_{*} \approx \msol{14}$) when the H~II region was around $650\unit{au}$ in extent. Because the initial reservoir was the smallest, there was less infalling gas and momentum to apply ram pressure compared to the higher mass cases. 

500-1.5-RAD demonstrated the most distinct trapped phase, remaining stalled with $r_\text{HII} \approx 717\unit{au}$ from $154\unit{kyr}$ ($M_{*} \approx \msol{18}$) to $225\unit{kyr}$ ($M_{*} \approx \msol{35}$). Note that during the same time frame, $r_\text{binding} \approx 73\unit{au}$, so the influence of the (proto)stellar gravity on the H~II region had been well surpassed, leaving ram pressure to dominate.

500-1.5-NORAD and 1000-1.5-NORAD also exhibited ram pressure trapping, despite simultaneous gravitational trapping which constrained the H~II region within $r_\text{binding}$. In 500-1.5-NORAD, after the onset of photoionization expansion, the H~II region rapidly expanded to $r_\text{binding}$. At the H~II region's maximum extent on the central axis, it remained at $r_\text{binding}$ for the duration of the simulation. On the central axis, however, the H~II region expanded to $r_\text{binding}$ but then experienced an oscillatory phase when the (proto)star was approximately $\msol{25}$. 

Similarly, in 1000-1.5-NORAD (the blue line in rightmost panel of Figure \ref{fig:rad_pressure_comparison}) at the onset of photoionization expansion, the ionized gas boundary quickly approached $r_\text{binding}$. However, the boundary at the central axis was just as quickly pushed back down from $45\unit{au}$ to $3.5\unit{au}$, nearly the same level as before the initial burst of expansion. The boundary then slowly climbed back up to $r_\text{binding}$, from $96$ to $158\unit{kyr}$. This long staggered climb contrasts clearly against 500-1.5-RAD or 1000-1.5-RAD as well as 250-1.5-NORAD, in which smooth and rapid expansion immediately reached or surpassed $r_\text{binding}$.

Given that ram pressure is the only physical process other than gravity directed towards the origin, the fact that $r_\text{HII}$ was pushed well below $r_\text{binding}$ on the central axis instead of being kept at $r_\text{binding}$ indicates that ram pressure from the infalling gas along the central axis must also have pushed the boundary back towards the (proto)star. Furthermore, the oscillatory nature of these stalled phases is well in agreement with observational studies such as \citet{galvan2008time}, which describes an HCH~II region unexpectedly contracting over very short timescales and proposes gravitational trapping and accretion as possible causes.

\subsection{Lifetimes}
We define the HCH~II lifetime as the total time from when $r_\text{HII}$ is $100\unit{au}$, to when it crosses the hypercompact boundary at $2000\unit{au}$ ($0.01\unit{pc}$). The UCH~II lifetime includes the HCH~II lifetime, and thus is the total time required to progress from $100\unit{au}$ to $20{,}000\unit{au}$ ($0.1\unit{pc}$).

See Table \ref{tab:phase_lifetimes} for the HCH~II and UCH~II lifetimes of all simulations. In the cases with radiation pressure and the density profile with $\beta_\rho = -1.5$, the UCH~II lifetimes were $101.3\unit{kyr}$, $164.2\unit{kyr}$, and $160.5\unit{kyr}$ for the $\msol{250}$, $\msol{500}$, and $\msol{1000}$ cases respectively. These lifetimes on the order of $10^5\unit{years}$ fit within the observational lifetimes suggested by works such as \citet{mottram2011rms}. In case 250-1.5-RAD, the HCH~II lifetime was extended both by an oscillatory gravitationally trapped phase and a brief ram pressure trapped phase. 500-1.5-RAD and 1000-1.5-RAD did not experience notable gravitational trapping but experienced extended stalling via ram pressure. 500-1.5-RAD experienced most trapping in the hypercompact phase, while 1000-1.5-RAD experienced growth staggered by short trapped phases throughout both the hyper and ultracompact phases. All cases with $\beta_\rho = -1.5$ also displayed periods of contraction; not only did stalling slow down H~II region growth, but the boundary of the ionized gas was actually forced back. See case 1000-1.5-RAD in the right panel of Figure \ref{fig:rad_pressure_comparison} at $\msol{40}$ ($122\unit{kyr}$) for an example of $r_\text{HII}$ decreasing due to ram pressure.

The cases without radiation pressure remained gravitationally trapped within the HCH~II phase for the entirety or the vast majority of the simulation time. The cases with the initial density profile of $\beta_\rho = -2.0$, with initial mass concentrated closer to the star, experienced rapid and uninterrupted expansion resulting in much shorter lifetimes on the order of $10^4\unit{years}$. 
\begin{table}[tb]
    \caption{Lifetimes of hypercompact (HC) and ultracompact (UC) phases for all simulations, measured from when the H~II region reached $100\unit{au}$ in size. The ``UC'' column is thus the time spent in both the hypercompact and ultracompact phases before exiting into the compact phase. * indicates that the given threshold (e.g. passing from hypercompact to ultracompact) was never reached, so the lifetime of that phase was indefinite since it did not end.}
    \label{tab:phase_lifetimes}
    \centering
    \begin{tabular}{r r r r r r}
        \hline\hline
        \# & $M_\mathrm{core} \mbox{ [M}_\odot]$ & $\beta_\rho$ & $P_{\mathrm{rad}}$ & HC [$
        \unit{kyr}$] & UC [$\unit{kyr}$] \\
        \hline
        0 & 250 & -1.5 & On & 69.7 & 101.3\\
        1 & 250 & -2.0 & On & 17.3 & 52.8\\
        2 & 500 & -1.5 & On & 107.4 & 164.2\\
        3 & 500 & -2.0 & On & 16.4 & 43.9\\
        4 & 1000 & -1.5 & On & 54.0 & 160.5\\
        5 & 250 & -1.5 & Off & * & *\\
        6 & 500 & -1.5 & Off & 221.0 & 235.4\\
        7 & 1000 & -1.5 & Off & 152.0 & *\\
    \hline
    \end{tabular}
\end{table}

\subsection{Morphology and centrifugal force}
We describe case 500-1.5-RAD here as a representative example of the morphological evolution seen in all simulations with radiation pressure. The gravitational collapse of the initial gas cloud leads to the formation of an optically thick, rotating accretion disk on scales of hundreds of au. The formation of this disk impedes infall along the midplane, shaping the environment into which the H~II region later expands. The presence of this dense, optically thick disk is the primary reason the H~II region assumes its characteristic ``butterfly-wing'' morphology. Unable to expand through the disk, the ionized gas is instead channeled into two lobes perpendicular to it, creating the observed shape.

\begin{figure}[htbp]
    \centering
    \begin{subfigure}{0.41\textwidth}
        \includegraphics[width=\textwidth]{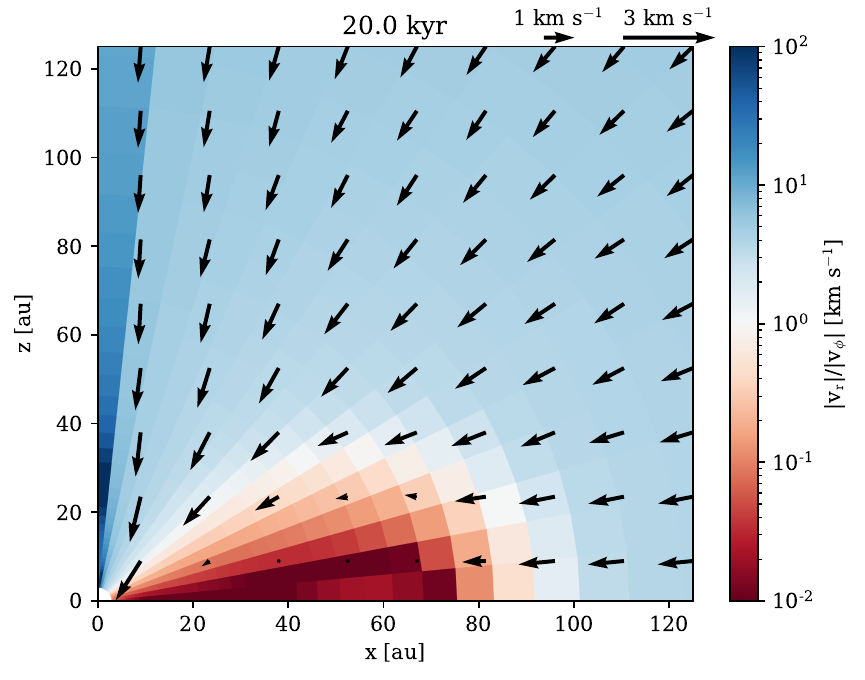} 
    \end{subfigure}\\
    \begin{subfigure}{0.42\textwidth}
        \includegraphics[width=\textwidth]{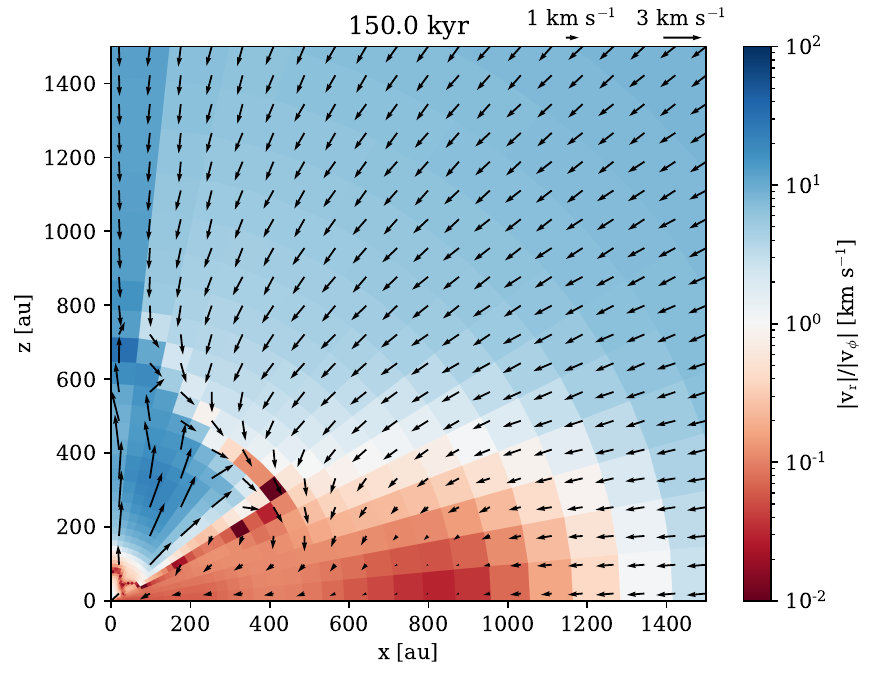} 
    \end{subfigure}
    \caption{Velocity before and after H~II region expansion began in run \#2. Background: ratio between radial and azimuthal velocity. Arrows indicate the velocity field in the xz plane. Note the presence of the torus, shown in red, where the magnitude of azimuthal velocity exceeds that of radial velocity, demonstrating the dominance of centrifugal force in particular regions on small scales relative to the whole cloud.}
    \label{fig:disk_velocity_field}
\end{figure}

Torus formation ($|v_r| < |v_{\phi}|$) began very early (around $300\unit{years}$) at the inner simulation domain boundary. Over the next $30\unit{kyr}$ the torus grew to around $100\unit{au}$. The top panel of Figure \ref{fig:disk_velocity_field} shows the state of torus at this time, before the onset of photoionization. By $50\unit{kyr}$, the torus had become an accretion disk ($v_{\phi} = v_\mathrm{Kepler}$) with an extent of $300\unit{au}$. The disk continued to grow until $100\unit{kyr}$, reaching a maximum extent of $1000\unit{au}$.

\begin{figure*}[tbp]
    \centering
    \begin{subfigure}{0.23\textwidth}
        \includegraphics[width=\textwidth]{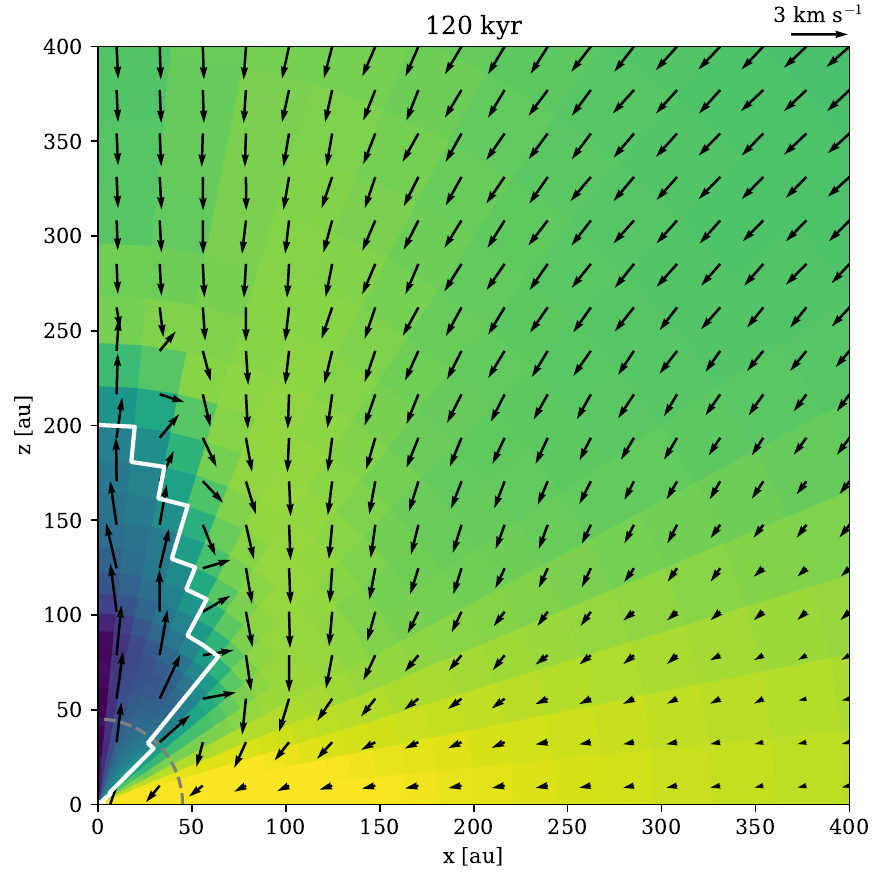}
    \end{subfigure}
    \begin{subfigure}{0.23\textwidth}
        \includegraphics[width=\textwidth]{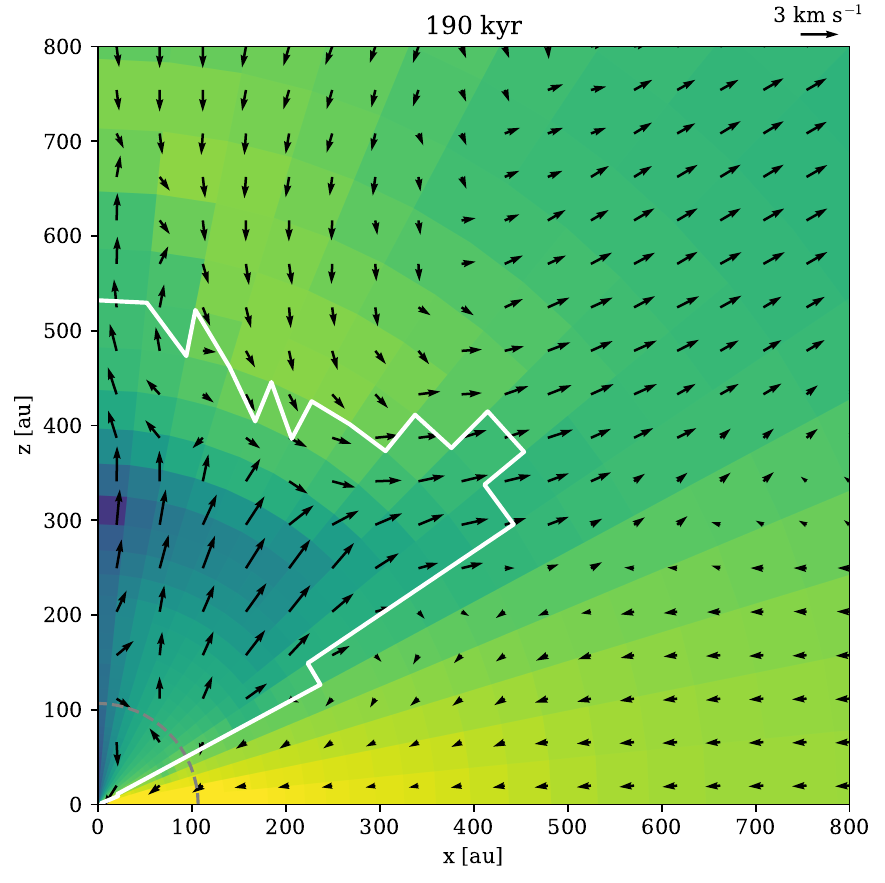}
    \end{subfigure}
    \begin{subfigure}{0.23\textwidth}
        \includegraphics[width=\textwidth]{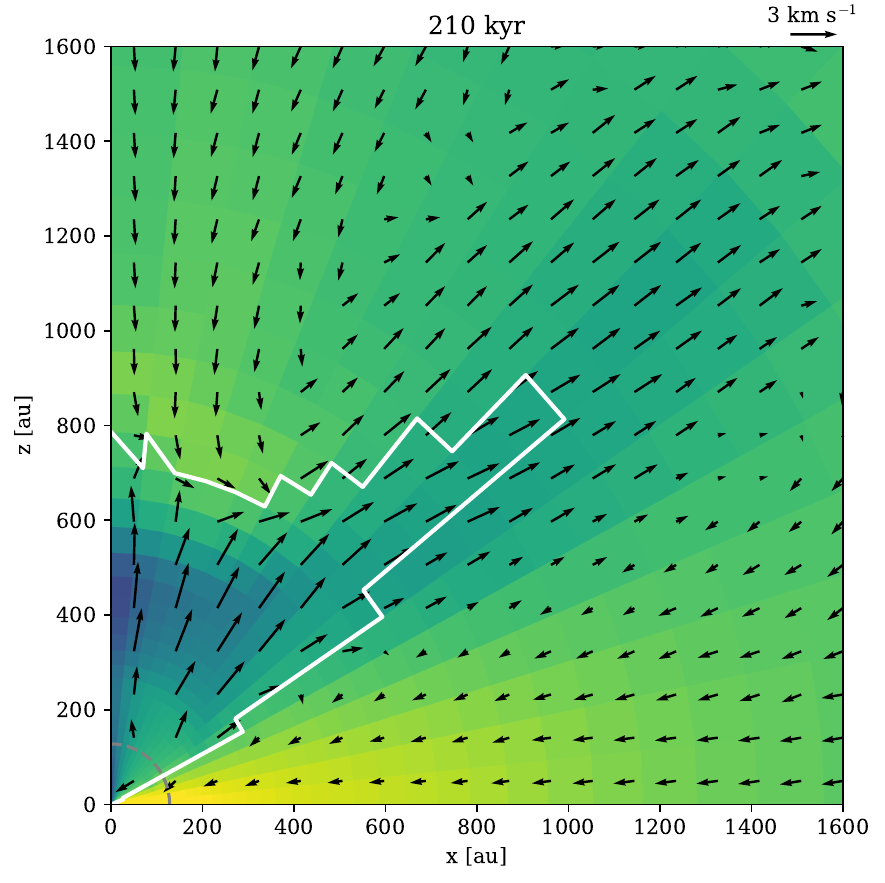}
    \end{subfigure}
    \begin{subfigure}{0.23\textwidth}
        \includegraphics[width=\textwidth]{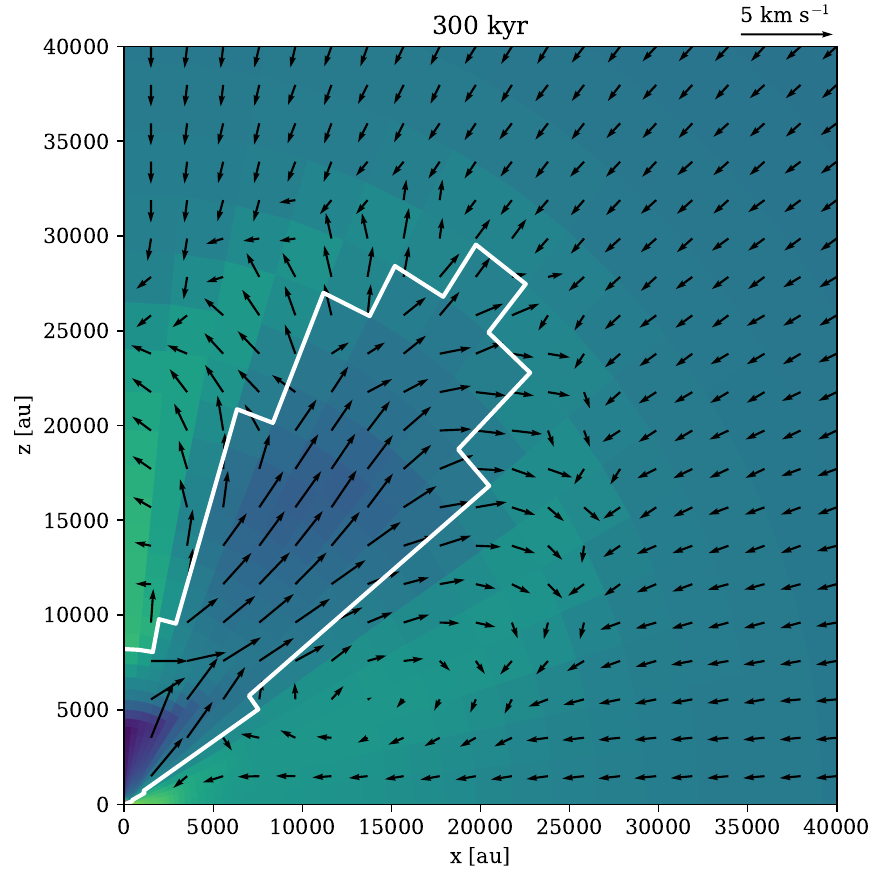}
    \end{subfigure}
    \begin{subfigure}{0.055\textwidth}
        \includegraphics[width=\textwidth]{morph_example/rho_butterfly_colorbar.pdf}
    \end{subfigure}
    \caption{Morphology of the H~II region during the early and middle stages of growth in run \#2. The white contour line indicates the boundary of the H~II region. Note that the scale increases with each panel. The grey dashed line is the gravitational binding radius.}
    \label{fig:centrifugal_morphology_quiver_early}
\end{figure*}


The optically thick interior of the accretion disk  
prevented ionizing radiation from entering the disk region. The disk thus acted as a boundary for the H~II region along the midplane. Note the boundary formed where the H~II region expanded outwards along the top edge of the disk, as seen in the bottom panel of Figure \ref{fig:disk_velocity_field}. Also note the flow of in-falling material as it collided with the H~II region boundary and was then diverted into the disk, applying ram pressure to the outer edge of the ionized region.

The H~II region itself underwent interesting morphological changes as it grew. Figure \ref{fig:centrifugal_morphology_quiver_early} shows the morphology of the H~II region at time-points from $120\unit{kyr}$ to $300\unit{kyr}$, with the presence of the disk visibly dictating which direction the ionized region could expand. Note that the flow of gas from the envelope into the accretion disk constrains the H~II region along the midplane at radii even beyond the edge of the disk. From $120\unit{kyr}$ to $190\unit{kyr}$, the H~II region expanded as a lobe around the central axis. At $190\unit{kyr}$, this lobe had broadened and a protrusion began to form along the side of the lobe adjacent to the disk, pushed outwards by thermal expansion pressure, radiation pressure, and centrifugal force. At $210\unit{kyr}$, the protrusion had become much more pronounced, and by $300\unit{kyr}$, the lobe had entirely separated from the central axis, forming the distinct wing structure as seen in 2D. The wing structure is maintained until the H~II region reaches the edge of the simulation domain. This shape also demonstrates the difference between $r_\text{HII}$ and $r_\text{HII, z}$, as the ``wing-tip'' represents the maximum extent of the H~II region away from the origin ($r_\text{HII}$), while the extent of the ionized region along the central axis ($r_\text{HII, z}$) lags behind (see Figure \ref{fig:butterfly_morphology} where this distinction is labeled).

Throughout our simulations, this morphology and structure is clearly visible in the neighborhood of the star due to the assumed axial symmetry and idealized initial conditions, particularly the lack of initial turbulence. The effect would likely be less pronounced in nature. Nonetheless, these morphologies provide the opportunity to analyse the influence of centrifugal force on the growth of the H~II region, by comparing the H~II region extent along the z-direction (where the centrifugal force is zero) with its maximum extent relative to the origin.

\subsection{Initial reservoir mass and density profile}
As the initial reservoir mass increased, the onset of photoionization occurred earlier. Each mass case exhibited similar phases of cloud collapse, disk formation, H~II region formation, and stellar evolution, but each step occurred more quickly in the $\msol{1000}$ case than in the $\msol{500}$ case, than in the $\msol{250}$ case. 

\begin{figure*}
    \centering
    \includegraphics[width=\textwidth]{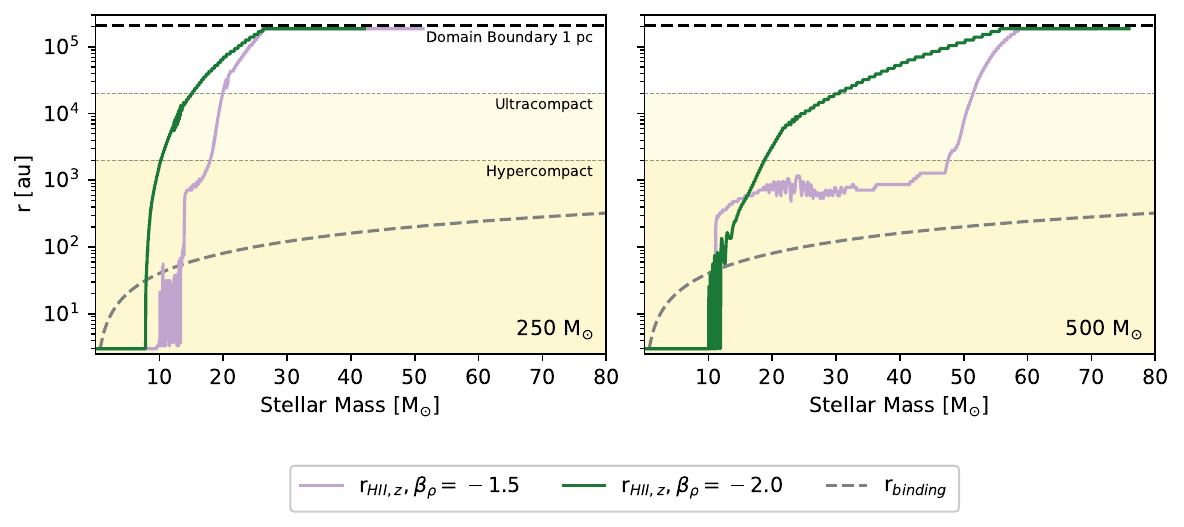}
    \caption{H~II region extent in the $\msol{250}$ (left panel) and $\msol{500}$ (right panel) cases, comparing initial density profiles of $\beta_{\rho} = -1.5$ (lavender: runs \#0 and \#2) and $\beta_{\rho} = -2.0$ (green: runs \#1 and \#3).}
    \label{fig:density_profile_comparison}
\end{figure*}

In the cases with the density profile where $\beta_\rho = -2.0$, meaning that the initial mass was distributed closer to the origin, gravitational collapse proceeded much more rapidly than in the corresponding mass cases where $\beta_\rho = -1.5$. 
Additionally, the initially lower mass on large scales resulted in lower ram pressure against the H~II region growth later on in evolution.
No gravitational or ram pressure trapping occurred in the $\msol{250}$ case, and only some brief oscillations occurred due to gravitational trapping in the $\msol{500}$ case. Note the quicker, uninterrupted expansion of the $\beta_\rho = -2.0$ cases as shown in Figure \ref{fig:density_profile_comparison}.

\subsection{Limitations and future work}
This work does not include magnetic fields, so magnetohydrodynamical (MHD) simulations would be a sensible progression. 
We also did not include a manual injection of protostellar outflows, purposefully, due to previous studies which showed no H~II trapping in the presence of outflows. Compare here with \citet{kuiper2018first}, which assumed that the (injected) protostellar outflow does not stop during the star-formation process. The injected outflows created large bipolar cavities, allowing for unhindered H~II region expansion.

The present study considers the opposite limit, that is, the absence of protostellar outflows. As a result of this assumption, we have obtained plausible scenarios for the trapping of H~II regions. Given the observational evidence for both limits to the growth of H~II regions (see \ref{Section-Lifetime Problem}) and considering the formation of magnetically-driven jets \citep{moscadelli2022snapshot} before photoionization in nature, this may suggest that magnetically-driven jets and outflows have a limited lifetime \citep[see][for a discussion on how such outflows can be terminated by magnetic braking]{oliva2023modeling} so that they do not impact the eventual growth of H~II regions.

We propose that the entire H~II region lifetime may occur (with trapped phases) in systems which never experience outflows, or after magnetic-forces have died down and the launching of jets and outflows has ended. There could be enough time in the early lives of massive stars for both a magnetically dominated phase characterized by outflows/jets, followed by a phase of H~II region expansion. Alternatively, there might be systems for which magnetically-driven outflows are too weak or intermittent, leading to trapping. Future works should examine this question in more detail by including MHD modeling of the flow that allows for the self-consistent launching of protostellar outflows.

Line-driven stellar winds were also not included, which could have had an effect on accretion rates \citep[see][]{kee2019line}. Ideally, future simulations would also include line-driven feedback, impacting disk ablation and stellar wind feedback.

Naturally, these simulations are limited by being two-dimensional. We assume axial and midplane symmetry to approximate the 3D nature of the system. Observationally, asymmetric structures such as spiral arms, fragments, and clumps occur. The rotational symmetry of the disk is thus also assumed, which influences the overall morphology of the system. Future simulations would ideally simulate a full 3D domain with no assumptions on symmetry, though of course this would require significantly greater computational resources and time. See e.g.~\citet{geen2023energy} for one such 3D simulation, which utilized a much larger box-shaped grid. They included MHD and stellar winds, and focused primarily on later stages of massive star formation at larger scales than this work. 

\section{Conclusion}
We performed radiative-hydrodynamics simulations to explore the roles of (proto)stellar gravity, ram pressure from infalling gas, radiation pressure, and centrifugal forces in stalling and supporting early H~II region growth. The simulations exhibited the formation of an accretion disk and the evolution of an H~II region. 
We found that H~II region lifetimes in the hypercompact and ultracompact phases could be extended by both early gravitational trapping at small scales and later ram pressure trapping at larger scales.

In the absence of radiation pressure, the gravitationally trapped phase lasted indefinitely. These H~II regions were thus in the hypercompact phase for the duration of the simulation. 
On the contrary, in the presence of radiation pressure, both early gravitational trapping at small radii and later ram pressure trapping at large radii occurred. Together, gravitational trapping and ram pressure trapping significantly stalled the H~II region growth via phases in which minimal growth or even contraction occurred. These fluctuations in H~II region radius agree with observational reports which describe decreases in UC/HC H~II region size over short timescales, such as in \citet{galvan2008time}. The resulting lifetimes of the hypercompact and ultracompact phases are roughly in accordance with observations. This suggests that these physical mechanisms are key in lengthening the duration of the compact phases.

Furthermore, we tested two initial density profiles, one in which the reservoir mass was distributed with a higher density closer to the origin ($\beta_{\rho} = -2.0$) than the other ($\beta_{\rho} = -1.5$). The centrally-concentrated density profile resulted in a shorter collapse and H~II region growth timescale, with no trapping, skipping immediately past the compact phases.

We acknowledge the limitations set by the axial and midplane symmetry and suggest that future work would ideally be full 3D, and include magnetic fields.

\begin{acknowledgements}
RK acknowledges financial support via the Heisenberg Research Grant funded by the Deutsche Forschungsgemeinschaft (DFG, German Research Foundation) under grant no.~KU 2849/9, project no.~445783058.
\end{acknowledgements}

\bibliographystyle{aa} 
\bibliography{BibFile}

\onecolumn
\renewcommand{\thefigure}{A\arabic{figure}}
\setcounter{figure}{0}
\section*{Appendix}
\FloatBarrier
\begin{figure*}[h]
    \centering
    \begin{subfigure}{0.375\textwidth} 
        \includegraphics[width=\textwidth]{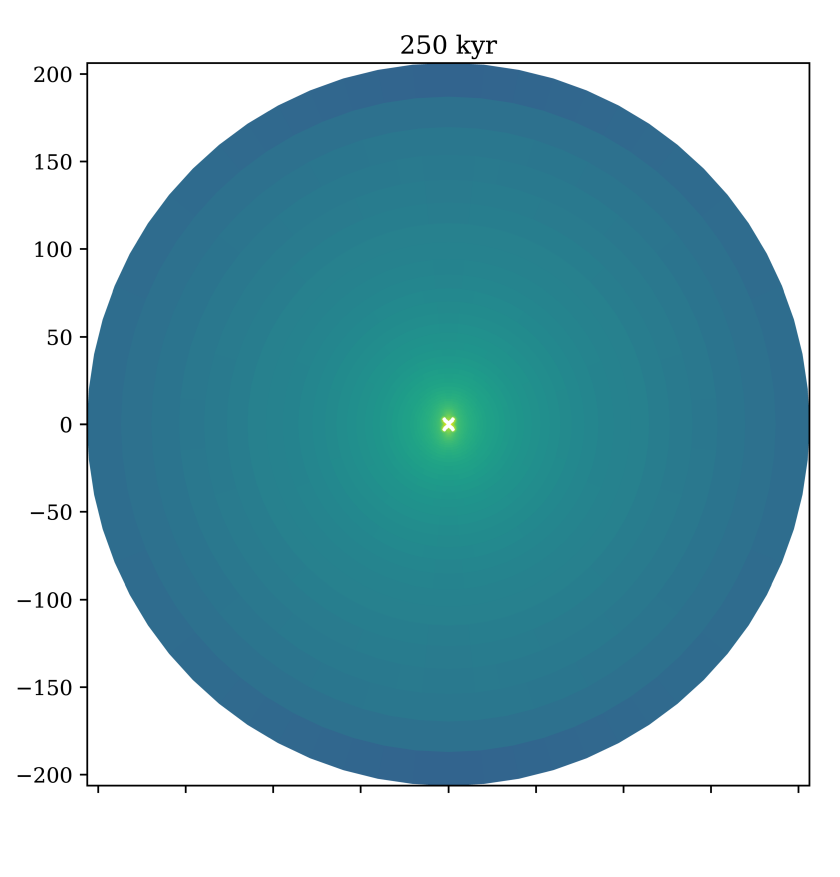}
    \end{subfigure}
    \begin{subfigure}{0.37\textwidth} 
        \includegraphics[width=\textwidth]{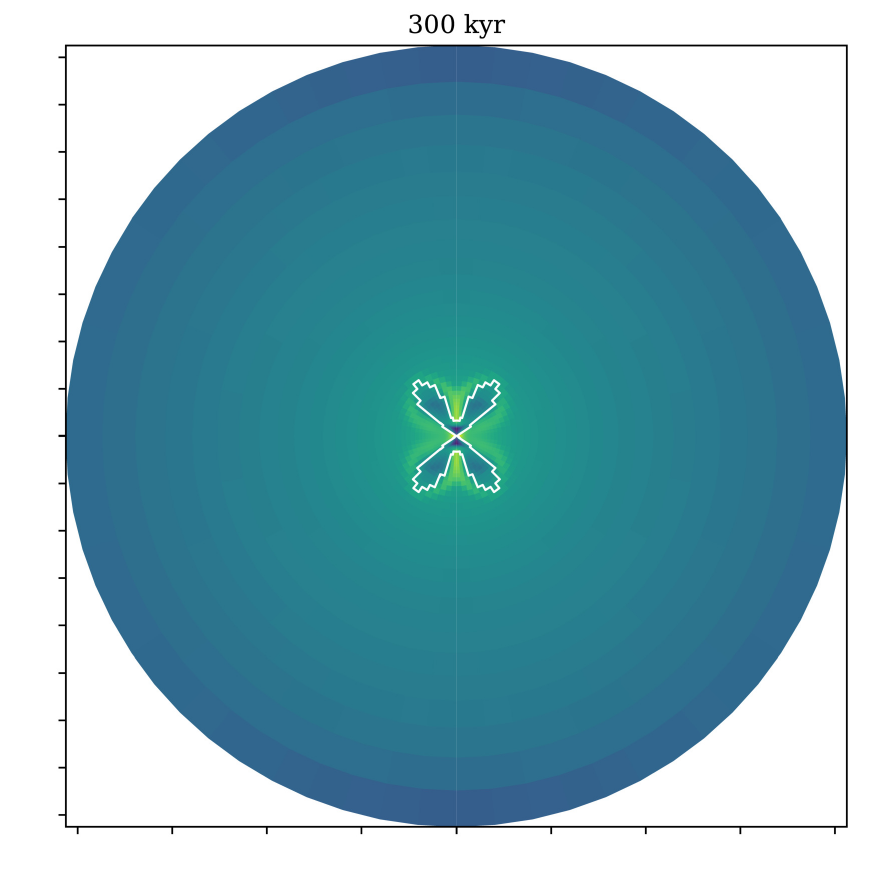}
    \end{subfigure}
    
    \vspace{-1em}
    \begin{subfigure}{0.375\textwidth}
        \includegraphics[width=\textwidth]{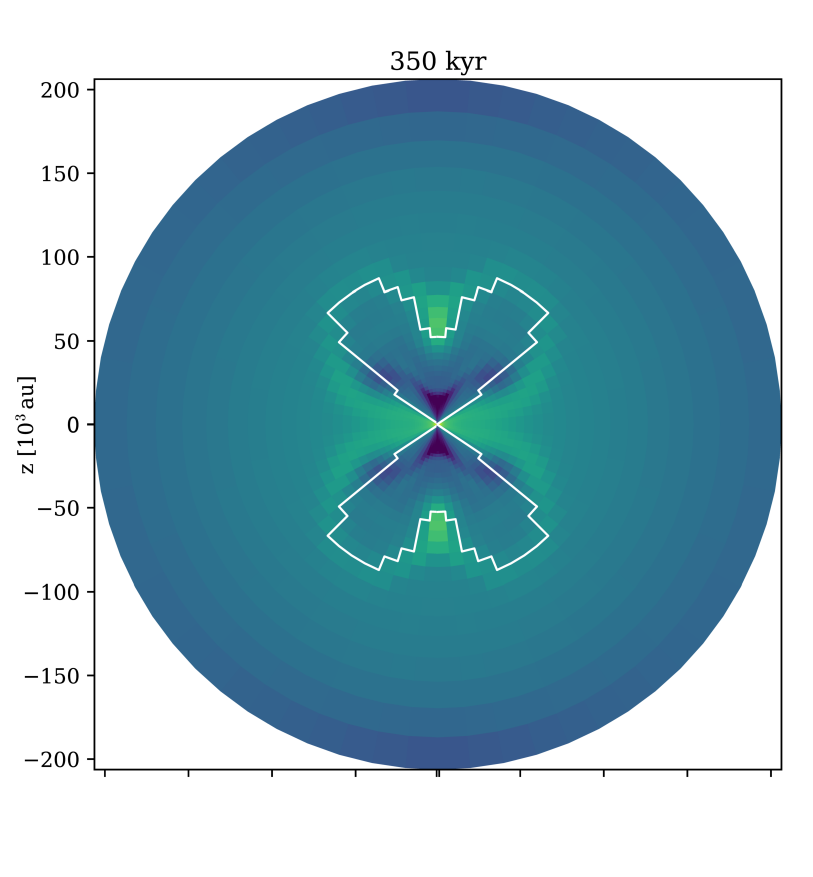}
    \end{subfigure}
    \begin{subfigure}{0.37\textwidth}
        \includegraphics[width=\textwidth]{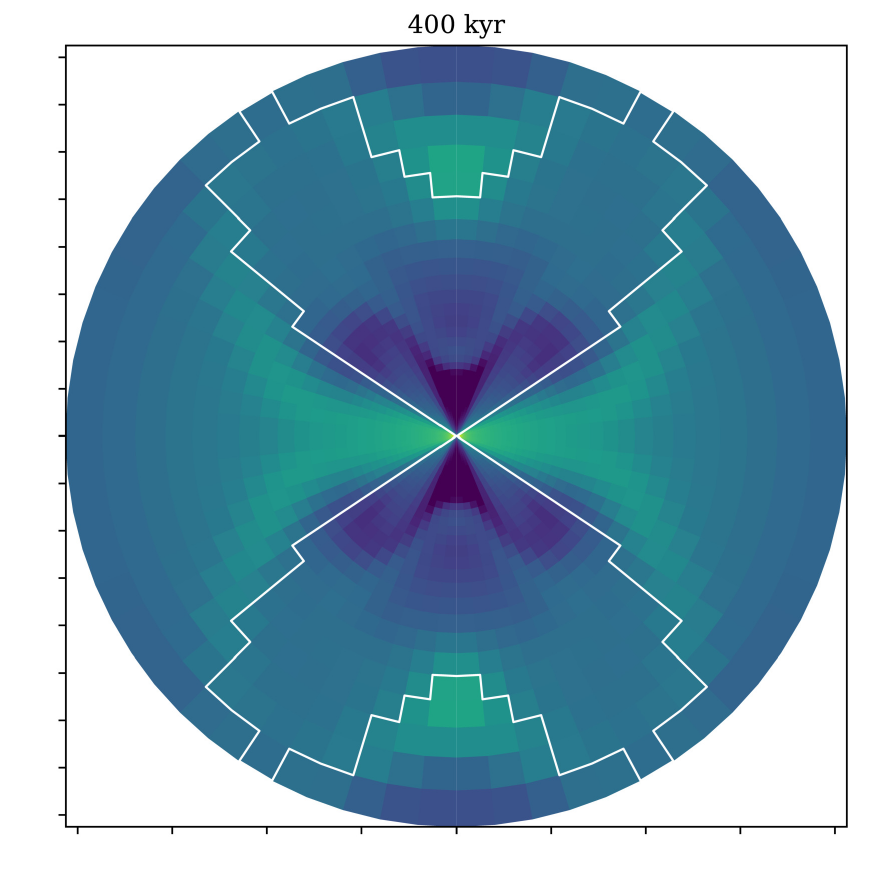}
    \end{subfigure}
    
    \vspace{-1em}
    \begin{subfigure}{0.375\textwidth}
        \includegraphics[width=\textwidth]{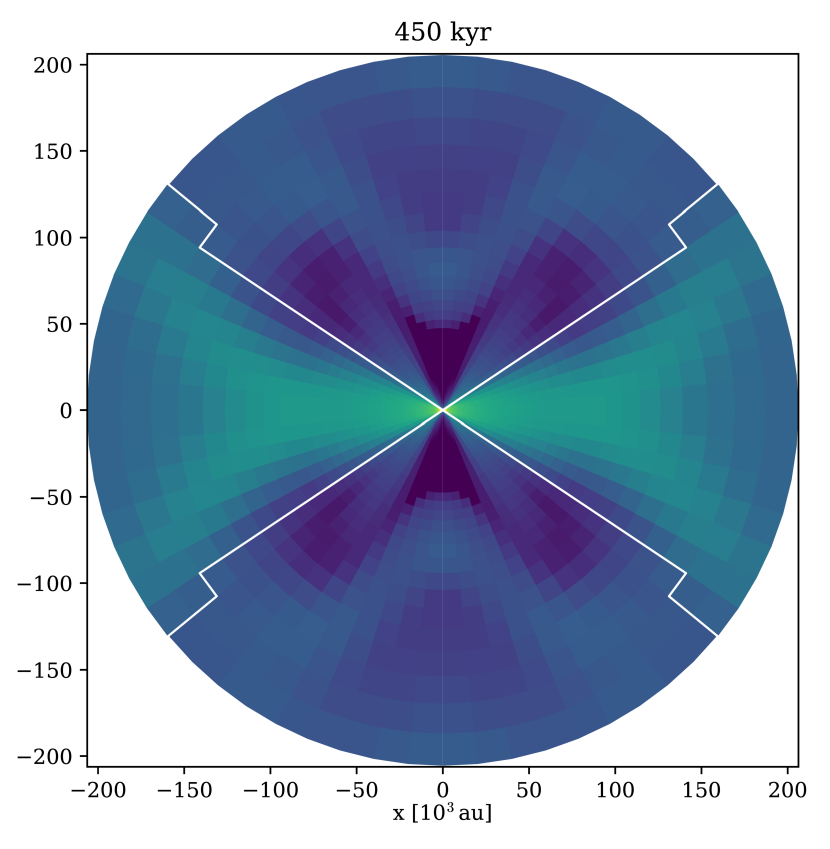} 
    \end{subfigure}
    \begin{subfigure}{0.37\textwidth}
        \includegraphics[width=\textwidth]{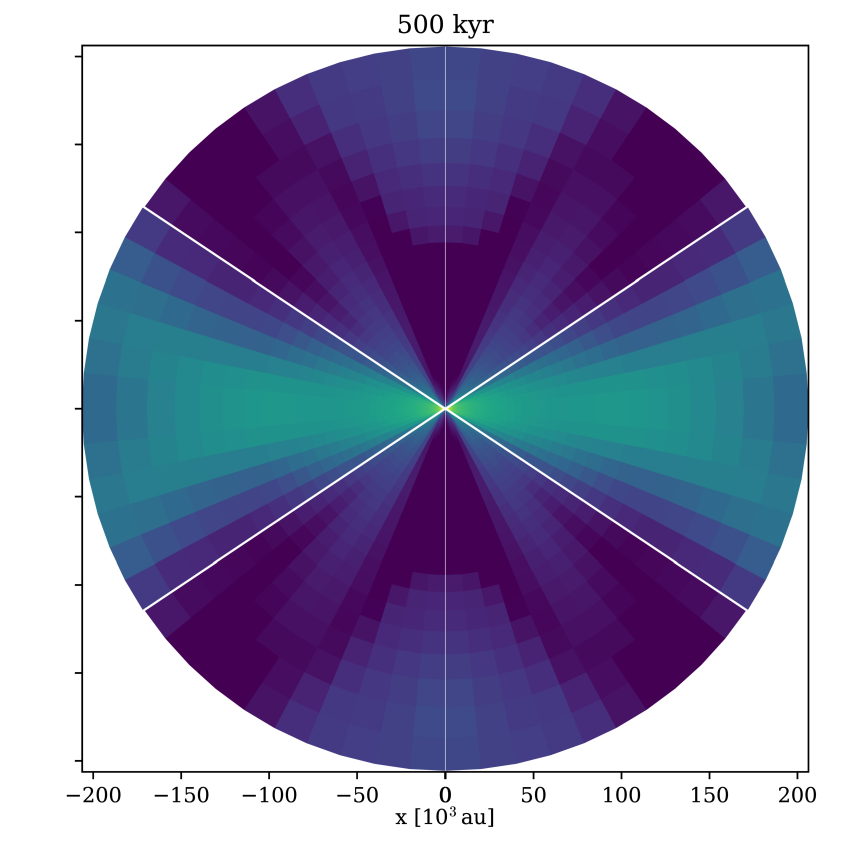}
    \end{subfigure}
    \caption{Morphology of the expanding H~II region at large scales with density in the background. The contour defined by the white line indicates the boundary of the H~II region, within which $>50\%$ of hydrogen was ionized. Snapshots were taken from run \#2 (500-1.5-RAD). The color-map is shared with Figure \ref{fig:butterfly_morphology}.}
    \label{fig:butterfly_expansion_snapshots}
\end{figure*}

\end{document}